\documentstyle[prb,aps,twocolumn]{revtex}
\makeatletter
\def\gsim{\compoundrel>\over\sim}

\def\compoundrel#1\over#2{\mathpalette\compoundreL{{#1}\over{#2}}}
\def\compoundreL#1#2{\compoundREL#1#2}
\def\compoundREL#1#2\over#3{\mathrel
   {\vcenter{\hbox{$\m@th\buildrel{#1#2}\over{#1#3}$}}}}
\makeatother
\input epsf.tex
\input psfig.tex
\begin{document}
\draft
\wideabs{
\title{Free vortex and vortex-pair lifetimes in
classical two-dimensional easy-plane magnets}
\author{D. A. Dimitrov}
\address{Los Alamos National Laboratory, Los Alamos, New Mexico 87545}
\author{G. M. Wysin}
\address{Department of Physics, 
Kansas State University, 
Manhattan, KS 66506-2601 }
\date{\today}
\maketitle
%
%
\begin{abstract} 
We report numerical simulation results for free-vortex lifetimes 
in the critical region of classical two-dimensional easy-plane ferro-
and antiferromagnets having three-component order parameters. 
The fluctuations in the vortex number density in a spin dynamics
simulation were used to estimate the lifetimes.
The observed lifetimes are of the same order of magnitude but smaller 
than the characteristic time-scale above which a phenomenological 
ideal vortex-gas theory that has been used to account for the central peak 
in the dynamic structure factor $S^{\alpha\alpha}({\bf q},\omega)$ 
is expected to be valid.
For strong anisotropy, where the vortices are in-plane, 
the free vortex lifetimes for ferromagnets and antiferromagnets 
are the same, while for weak anisotropy, where the vortices have
nonzero out-of-easy-plane components, the lifetimes in
antiferromagnets 
are smaller than in ferromagnets.
The dependence of the free-vortex and total vortex densities on the
size dependent correlation length in the critical region is examined.
We also determined the lifetimes of vortex-antivortex pairs for
$T = T_{\mbox{\scriptsize KT}}$ and well below
$T_{\mbox{\scriptsize KT}}$.
The observed time-scales are very short, and the observed pair
densities are extremely small.
These results suggest that pair creation and annihilation
are unlikely to 
play any role in the central peak in $S^{xx}({\bf q},\omega)$ 
observed in computer simulations for the ferromagnetic model for 
$T \leq T_{\mbox{\scriptsize KT}}$.
\end{abstract} 
} 

\narrowtext

\section{Introduction}

The interpretation of the central peak in the dynamic
structure factor $S^{\alpha\alpha}({\bf q},\omega)$ in
easy-plane layered ferromagnets (FM) and antiferromagnets (AFM)
is currently based on the phenomenological theory developed by
Mertens
{\it et al.}\cite{Mer87,Gou89,Vol92b,Vol91,Mer91,Vol92a,Iva94}
The central peak is observed for $T > T_{\mbox{\scriptsize KT}}$
($T_{\mbox{\scriptsize KT}}$ is the Kosterlitz-Thouless transition
temperature) and the theory accounts for it within the frame of
a dilute gas of free vortices effectively assuming
infinite free-vortex lifetime.
It is of importance to determine the free-vortex lifetime to
find out the interval of applicability of the theory as well as to
understand the time scale of the processes which account for the
finite lifetime.
We have developed a method to calculate the vortex lifetime
and we present here our results obtained from combined
cluster Monte Carlo (MC) and spin dynamics simulations.
We believe that these are the first reported
free-vortex
lifetimes in AFM for the system Hamiltonian we have considered.
The intensity and width of the central peak predicted by the 
dilute-vortex-gas theory depend on the free vortex density $N_v$
and the rms vortex velocity  $\bar{u}$. 
A free vortex is assumed to exist longer than a characteristic
observation time
\begin{equation}
\label{char_time}
1/\gamma = 2\xi / \sqrt{\pi}\bar{u},
\end{equation}
where $\xi$ is the correlation length.
It is assumed that the vortices have a Maxwellian velocity 
distribution.
If this theory is valid, it could, in principle, 
allow one to determine the average free vortex velocity $\bar{u}$ 
and the correlation length $\xi$ at a given temperature 
by experimentally measuring the shape of the central peak, 
knowing the strength of the Heisenberg exchange 
interaction for the magnetic material.
The FM and AFM static properties on a square lattice 
(or other bi-partite lattice) are identical (the spins on one of the
sublattices in the AFM are inverted when compared with the spins in 
the FM) while their dynamic properties differ.\cite{Iva95}
In particular, the out-of-plane
tilting of spins near the core of a vortex produces a nonzero topological
charge (or gyrovector\cite{Thiele73,Huber80}) for vortices in the FM model.
The gyrovector plays an important role in determining how the motion of
an individual vortex is influenced by effective fields due to neighboring
vortices, and it appears together with an effective mass in a collective
coordinate equation that describes the motion of a vortex center.\cite{Wys94}
However, the gyrovector is always zero for vortices in the AFM model, and,
the effective mass for vortices in the AFM model is much smaller than 
for vortices in the FM model.\cite{Wys96}
For these reasons we expect that the FM and AFM models could have quite 
different vortex lifetimes, and therefore contrast the lifetimes measured 
for the AFM model with those we have determined previously for the FM 
model.\cite{dad96}

\section{The Model}
The Hamiltonian of the model is
\begin{equation} 
\label{Hamiltonian}
H = -J\sum_{\langle i,j\rangle }(S_i^xS_j^x + S_i^yS_j^y + 
			\lambda S_i^zS_j^z),
\end{equation} 
where the classical spins $\mbox{\bf S}_i = (S^x_i,\ S^y_i,\ S^z_i)$
are vectors on the unit sphere $S^2$,
the sum is over nearest-neighbor
sites of a square lattice, the easy-plane anisotropy parameter $\lambda$
varies in the interval $0\leq\lambda < 1$, and 
$J > 0$ for FM and $J < 0$ for AFM.
We assume that $J$ has energy units, time is measured in units of
$\hbar/J$ and temperature in $J/k_B$; the spins are dimensionless.  

The static critical behavior of this model is well-described by the 
extensively studied classical XY model (the two-dimensional $O(2)$ model)
but because of the possibility for spin fluctuations in one more 
degree of freedom (the out-of-plane $S^z$ fluctuations) the properties
of the XY model will be observed at lower temperatures here.
Long-range order cannot exist for any nonzero temperature in these
models,\cite{MW66} but Berezinskii\cite{Ber70} showed 
the existence of a phase transition which was understood by Kosterlitz 
and Thouless\cite{KT73} to occur through unbinding of pairs of 
nonlinear topological excitations--vortices.
The critical properties of the XY model\cite{Kos73} as well as the 
low-temperature phase are of considerable
interest to the physics of low-dimensional magnetism because there are
a number of materials which are effectively considered as
two-dimensional magnetic systems\cite{Jon90} and the Hamiltonian
(\ref{Hamiltonian}) is the simplest model to study their properties.
These are layered crystals with intralayer exchange interaction much 
greater than their interlayer exchange (more than two or three orders of 
magnitude greater) and there are observed both
ferromagnets\cite{Jon72,Kir82,Ain87,Bra88}  (e.g. K$_2$CuF$_4$,
(CH$_3$NH$_3$)$_2$CuCl$_4$, (C$_2$H$_5$NH$_3$)$_2$CuCl$_4$,
Rb$_2$CrCl$_4$) and antiferromagnets\cite{Reg89,Ros90,Gav91}
(like BaNi$_2$(PO$_4$)$_2$, BaCo$_2$(AsO$_4$)$_2$).
The intralayer exchange interaction is more than $10^{-3}$ orders of
magnitude smaller than the intralayer exchange in the graphite-intercalated
CoCl$_2$ which has been studied in detail by
Wiesler {\it et al.}\cite{Wie94}
There are magnetic lipid monolayers,\cite{Pom84} e.g. the compound
Mn(C$_18$H$_35$O$_2$)$_2$, which are true two-dimensional magnets.

When the anisotropy parameter $\lambda$ is less than a critical value,
$\lambda_c$ which is lattice dependent,\cite{Wys88} only in-plane
static vortex spin configurations ($\left<S^z\right>=0$) are stable,
while for $\lambda > \lambda_c$ only static vortices with nonzero
out-of-plane ($\left<S^z\right>\neq 0$) spin components are stable.
The critical anisotropy parameter $\lambda_c$ has recently been
determined\cite{Wys97} with good precision and for the square lattice
$\lambda_c\approx 0.7034$.

\section{Dynamics}
The differences in dynamical properties of FM vs.\ AFM not only appear in the
vortex properties, but also in the spinwave-vortex interaction and
the spinwaves alone.
The spin-wave excitations have a single branch for FM while there are
two branches for AFM due to the two different spin sublattices.
The spin waves are either in-plane or out-of-plane depending on the 
value of $\lambda$.
Ivanov {\it et al.}\cite{IKW96} found a localized mode for the 
out-of-plane vortex in AFM which appears in the gap between the two
spin-wave branches.
Wysin {\it et al.}\cite{Wysin+97} have also shown that even for the
in-plane vortices in the AFM model, this localized mode is still present.
For the FM model, in contrast, only quasi-local spinwave modes appear 
on a vortex (where the $S^z$ component of the mode is localized near
the vortex core while the $S^x, S^y$ components of the mode are extended). 

In the ideal vortex-gas
theory,\cite{Mer87,Gou89,Vol92b,Vol91,Mer91,Vol92a,Iva94}
the motion of free vortices in a temperature interval just above
$T_{\mbox{\scriptsize KT}}$ is assumed ballistic and leads to central peaks 
in both in-plane $S^{xx}(q,\omega)$ ($S^{xx}=S^{yy}$ due to
the symmetry of the
Hamiltonian) and out-of-plane $S^{zz}(q,\omega)$
dynamic structure factors, where $S^{\alpha\alpha}(q,\omega)$
($\alpha = x, y, z$) is the space-time Fourier transformation of the
correlation function
$\langle S^\alpha(\mbox{\bf r}, t)S^\alpha(\mbox{\bf 0}, 0)\rangle $.
The central peak (CP) of the in-plane dynamic structure factor
is predicted to have a squared Lorentzian shape
\begin{equation}
S^{xx}(\mbox{\bf q}, \omega ) = {S^2\gamma^3 \xi^2 \over 2\pi
	\left[\omega^2 + \gamma^2( 1 + 
	(\xi q^*)^2)\right]^2},
\end{equation}
where $\gamma$ is defined in (\ref{char_time}).
The CP is located at $\mbox{\bf q}^* = (0,0)$
for FM ($\mbox{\bf q}^* \equiv \mbox{\bf q}$) and at the Bragg point
$\mbox{\bf K} = (\pi ,\pi)$ for AFM
($\mbox{\bf q}^* \equiv \mbox{\bf K} - \mbox{\bf q}$).
A weaker CP is predicted for the out-of-plane dynamic structure factor for
any $\lambda$,  due to correlations caused by vortex motion.
It is located at $\mbox{\bf q} = (0,0)$ for both FM and AFM.
Its intensity is proportional to the free vortex density $N_v$ and
has Gaussian shape
\begin{equation}
S^{zz}(q,\omega) = {N_v\bar{u}\over 32 f^2_\lambda J^2\sqrt{\pi}q}
	\exp\left(-\left(\omega/\bar{u}q\right)^2\right),
\end{equation}
where $f_\lambda = 1-\lambda$ for FM and $1+\lambda$ for AFM.
For $\lambda > \lambda_c$, the out-of-plane vortex spin asymptotic 
behavior is known $S^z(r)\sim \sqrt{r_v/r}\exp(-r/r_v)$, where
$r_v=\sqrt{\lambda/(1-\lambda)}/2$ is considered as the vortex core
radius and $r$ is the distance from the vortex center.
This asymptotic form is used to calculate $S^{zz}(q,\omega)$
for both FM\cite{Gou89} and AFM\cite{Vol92a,Iva94} leading to an 
additional CP {\em only} for $\lambda > \lambda_c$ with Gaussian shape
\begin{equation}
S^{zz}(q^*,\omega)\sim {N_v\bar{u}\over {q^*}^3}
	\exp\left(-\left(\omega/\bar{u}q^*\right)^2\right),
\end{equation}	
and again $\mbox{\bf q}^* \equiv \mbox{\bf q}$ for FM and
$\mbox{\bf q}^* \equiv \mbox{\bf K} - \mbox{\bf q}$ for AFM.
By measuring the width and the integrated intensity of the CP one can
determine $\bar{u}$ and $\xi$ and compare with independent theoretical
results.
We compare the characteristic times $1/\gamma$ obtained in this way with
our results on the vortex lifetime.
Thus, we expect that vortices in FM and AFM for $\lambda<\lambda_c$
have quite similar dynamics (and similar lifetimes),
which is reflected particularly in their 
similar in-plane correlations. 
However, the nonzero gyrovector for FM vortices is present for 
$\lambda > \lambda_c$.
This is an additional term in the vortex equation of motion for the
FM case, which is absent for AFMs.
Furthermore, the effective vortex mass of AFM vortices has been estimated
to be smaller than for FM vortices\cite{Wys96} only
for $\lambda > \lambda_c$. 
Because of these differences, we calculate the vortex lifetime for both
$\lambda < \lambda_c$ and for $\lambda > \lambda_c$, with the expectation
that larger differences should occur in the latter case.
In recent numerical simulations,\cite{Eve96} CP has been observed in
$S^{xx}(q,\omega)$ for $T\leq T_{\mbox{\scriptsize KT}}$
and $\lambda = 0$.
It is not clear currently what causes the CP for
$T\leq T_{\mbox{\scriptsize KT}}$.
For these temperatures, the dominant excitations 
are spin waves and the vortices are bound in pairs
with opposite vorticities.
The vortex pairs should only renormalize the shape of the spin-wave peaks
and should not lead to additional peaks as do the free vortices above
$T_{\mbox{\scriptsize KT}}$.
Nevertheless, we have calculated and checked
the lifetime of vortex pairs below
$T_{\mbox{\scriptsize KT}}$.
We found that the observed pair
lifetime is a very short time scale to be the relevant effect
for the observed CP at these temperatures.

%
%
\section{The Simulation}

We simulate a  system of classical spins (three-component unit
vectors) on a square lattice with periodic boundary conditions
and $L\times L$ number of sites, for sizes $L \leq 256$.
First, we run Monte Carlo (MC) simulations to obtain initial
equilibrium configurations (IC) and to calculate
the correlation length and vortex densities.
The obtained IC's are then evolved in time using a fourth order
Runge-Kutta method\cite{NR} to solve the Landau-Lifshitz
equations of motion
\begin{equation}
\label{em}
{d\mbox{\bf S}_i\over dt} = \mbox{\bf S}_i\times
	(-{\delta H\over \delta \mbox{\bf S}_i}),
\end{equation}
where
\begin{equation}
-{\delta H\over \delta \mbox{\bf S}_i} =
J \sum_{\langle j(i)\rangle }(S_j^x\mbox{\bf \^{x}}
 + S_j^y\mbox{\bf \^{y}} + \lambda S_j^z\mbox{\bf \^{z}}),
\end{equation}	
{\bf \^{x}}, {\bf \^{y}}, and {\bf \^{z}} are unit vectors along 
the coordinate system axes.
The sum is over the nearest neighbor sites of the site $i$.
We used a combination of Wolff's one cluster algorithm\cite{Wol89}
(for updating of the in-plane spin components $S_i^x$ and $S_i^y$),
Metropolis' algorithm,\cite{Met53} and the over-relaxed
algorithm\cite{Cre87} to update the spins in the MC part of the
simulations.
Each Monte Carlo step (MCS) through the system consists of three
single-cluster updates, three Metropolis sweeps, and three over-relaxed
sweeps.
The use of Wolff's algorithm is {\em essential} for reducing critical
slowing down.
The over-relaxed move on a classical spin consists of rotating the
spin around the direction of the local field due to its nearest 
neighbor spins (assuming no external magnetic field) at an angle of 
$\pi$, i.e. if
$\mbox{\bf \^{h}} = \mbox{\bf h}_{loc}/|\mbox{\bf h}_{loc}|$, then
\begin{equation}
\mbox{\bf s}_{new} =
	2 (\mbox{\bf \^{h}}\cdot\mbox{\bf s}_{old})\mbox{\bf \^{h}}
	-\mbox{\bf s}_{old}. 
\end{equation}	
The move is microcanonical and it also reduces critical
slowing down even though it is a local move. 
The Metropolis algorithm is needed to satisfy ergodicity of the MCS for 
the Hamiltonian we study.
The first 1000 to 5000 MCS were used for equilibration of the system
and IC were written after each bin consisting of 2000 to 10000 MCS.
Between 25 to 100 IC were produced for the spin dynamics simulations.
The calculation of the correlation length and vortex density is from
averages over 20 or 25 bins with measurements after each MCS and all 
estimated statistical errors are equal to the standard deviation of the
bin averages. 
The correlation length is determined by measuring\cite{Coo82,Kim9+,CTV95}
\begin{equation}
\xi_L = {1\over q} \sqrt{G(\mbox{\bf 0})/G(\mbox{\bf q}) - 1},
\label{XiL}
\end{equation}	
\begin{equation}
\label{Gk}
G(\mbox{\bf q}) = {1\over L^2} \left<\left| \sum_{\mbox{\bf r}_i}
\exp(i\mbox{\bf q}\cdot\mbox{\bf r}_i)\mbox{\bf S}_i^\perp\right|^2\right>,
\end{equation}
where $\mbox{\bf S}_i^\perp = (S_i^x,\, S_i^y)$ is the in-plane spin vector.
The wave-vector 
$\mbox{\bf q} = (2\pi/L,\ 0)$ is used,
and $\mbox{\bf r}_i$ is the radius-vector of the $i$th lattice site.
The correlation length $\xi_L$ obtained from Eq.\ (\ref{XiL}) approaches
the exact value as $O(q^4)$ approaches zero when increasing the linear
size of the system L.\cite{Kim9+,Man91}
The finite-size effects for the classical XY model 
are small,\cite{Wol89a} as a rule of 
thumb, when $L/\xi_L > 6$.

Another method used to determine the correlation
length\cite{Wol89a,TC79,Gupta88,Gup92} in the XY model from
simulations on a lattice is from a fit of
the zero spatial momentum two-point correlation function to
\begin{equation}
\Gamma(y_i)\equiv\left<\sum_{x_i,x_j}
	\mbox{\bf S}_{(x_i,y_i)}^\perp\cdot
	\mbox{\bf S}_{(x_j,0)}^\perp\right> =
	C\cosh\left(\left(L/2 - y_i\right)/\xi\right),
\end{equation}
the summation is over the x-coordinates of the lattice sites.
The advantage of using Eq.~(\ref{XiL}) to calculate the correlation 
length is that it does not require fitting.
We have also calculated $\Gamma(y_i)$ for several temperature values
and the comparison between the correlation lengths obtained from
Eq.~(\ref{XiL}) and from\cite{Gup92}
\begin{equation}
\xi(y_i) =
\ln\left(\Gamma\left(y_i\right)/\Gamma\left(y_i - 1\right)\right),
\label{XiG}
\end{equation}
at saturation (neglecting the contributions due to the periodic boundary
conditions in Eq.~(\ref{XiG})), shows that the two results agree within
2.5\%.
%

%
%
\section{Results}

\subsection{Free Vortices and Correlation Length}
We consider in this section how free vortices are determined and the
associated difficulties when applied to simulations on a lattice.
For the calculation of the vortex lifetime we use a method to
locate ``free'' vortices\cite{dad96} which is not directly related to the
correlation length and therefore it is {\em very important} to understand
how it relates to a method directly using $\xi(T)$ as the length
scale to distinguish free from bound vortices.
This requires the calculation of the correlation length for different
values of $T$ and $\lambda$ which we present below.
The results for the correlation length by themselves should be of
general interest, in particular, the values of $\xi(T)$ for
$\lambda = 0.9$ may not be reported elsewhere.
According to the  Kosterlitz and Thouless theory,\cite{KT73} at
$T_{\mbox{\scriptsize KT}}$ vortex-antivortex pairs start to unbind 
and free vortices are formed for $T > T_{\mbox{\scriptsize KT}}$.
Treating the core radius of a vortex as a variational parameter and
minimizing the vortex energy with respect to it, one can show that the
vortex core radius is proportional to the correlation length $\xi(T)$.
Thus, $\xi(T)$ sets a length scale below which vortex-antivortex
pairs can still be considered bound while above it the vortices are free.
The average free vortex density becomes
\begin{equation}
\label{nv}
n_v(T)\sim 1/\xi^2(T),
\end{equation}	
and there are arguments\cite{Iva95} that the exact dependence is
$n_v(T) = 1/(4\xi^2(T))$.
This suggests the following approach in a search
for free vortices on a lattice.
A vortex is considered free, provided 
the minimum distance from its center to a nearest neighbor vortex is
greater than $\xi(T)$. 
If the distance is less than $\xi(T)$, the corresponding two vortices are
marked as bound.
There are two related problems which render this approach too CPU intensive
currently.
First, the simulations are on a lattice and a vortex center can be
at any point inside a plaquette ($1\times 1$ square on the lattice
with spins located at its corners), where the vortex has been found.
We use Tobochnik and Chester\cite{TC79} method, which accounts correctly for
measuring spin in-plane angles $\mbox{mod}(2\pi)$, to
determine which plaquettes contain vortices.
Fitting to the known vortex solution for the spin
$\mbox{\bf S}_i = (\cos\theta_i\cos\phi_i,\ \cos\theta_i\sin\phi_i,
\ \sin\theta_i)$, with in-plane angle 
\begin{equation}
\phi_i = q\tan^{-1}\left({y_i-y_v\over x_i-x_v}\right),
\end{equation}	
where $q=\pm 1$ is the vorticity and $(x_v,\ y_v)$ are the coordinates of the
center of the vortex (treated as fitting parameters), allows 
determination of the vortex center location but creates the second problem.
The vortex positions and their free/bound status have to be determined at
each time step during the spin dynamics part of the simulations in order to 
determine the free vortex lifetime.
The fitting procedure for precise vortex positioning increases the CPU time 
by more than one order of magnitude (and close to two) which makes the 
simulations impractically long for large enough lattice sizes
(in order to have 
negligible finite size effects).
In addition, using the correlation length to determine the free
vortices for $\xi(T) \gsim 5.0$ gives approximately no free vortices.
For temperatures such that $\xi(T) \gsim 5.0$ the vortices are still
distributed in clusters and the correlation length is usually greater than
the minimum distance between nearest neighbors of vortices.

\begin{figure}
\mbox{\hspace{.3in}\psfig{figure=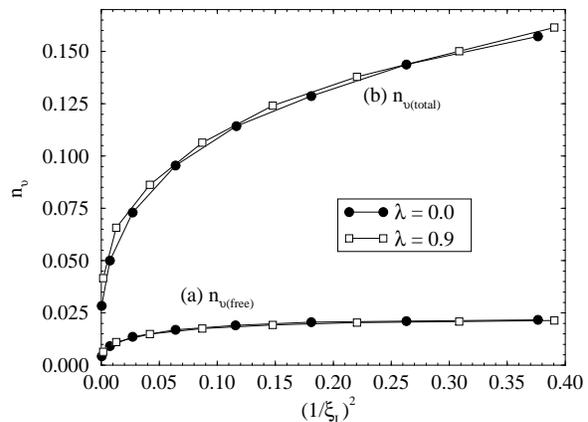,height=2.75in,angle=-90.}}
\caption{
  The total and free vortex densities  
  vs $1/\xi_L^2$ for $\lambda = 0.0$ and $0.9$. The errors are smaller than  
  the size of the data point symbols.}
\label{fig1} 
\end{figure}

Therefore, we followed our previuos approach\cite{dad96} for locating
free vortices.
We take the vortex positions as the centers of the plaquettes, which
eliminates the most CPU intensive task of precisely fitting their
positions. 
We also use a fixed length scale to determine the
free/bound vortex status in the temperature interval we study.
This length scale is set equal to the next nearest neighbor distance,
i.e. $\sqrt{2}$ lattice constants on the square lattice.
A bound vortex will have a nearest neighbor vortex at a distance of 
one lattice constant or $\sqrt{2}$ while a free vortex may have its 
nearest neighbor vortex at a distance greater or equal to 2 lattice constants.
This method to determine the free vortices has already been used by Gupta and 
Baillie\cite{Gup92} to measure the vortex density.
We determined the correlation lengths and vortex densities for the
temperature intervals we studied and two values of the easy-plane
anisotropy parameter ($\lambda = 0.0$ and $\lambda = 0.9$) in order to 
estimate the applicability of the fixed-length ($\sqrt{2}$)
approach to determine the free vortices.
With the critical temperatures estimated as 
$T_{\mbox{\scriptsize KT}}(\lambda = 0.0)\approx 0.70$ and 
$T_{\mbox{\scriptsize KT}}(\lambda = 0.9)\approx 0.63$, 
we studied the range
$0.75\leq T \leq 1.1$ for $\lambda = 0.0$ and
$0.7\leq T \leq 1.05$ for $\lambda = 0.9$ at intervals of
$\Delta T = 0.05$.
The critical temperature was estimated using the reduced forth-order
cumulant.\cite{Privman90}
For $T$ lower than these intervals, the free-vortex density is too low
to obtain good statistics.
The results for the correlation length are presented in Table~\ref{table1}
for $\lambda = 0.0$ and $\lambda = 0.9$.
We expect that these values of $\xi_L$ have reached saturation and
should be approximately equal to the 
values of $\xi$ in the thermodynamic limit with the possible exception of
the data point at $T = 0.75$ for $\lambda = 0.0$ and at $T = 0.7$ for
$lambda = 0.9$.
These values of the correlation length suggest that the use of
fixed length equal to $\sqrt{2}$ to determine the free 
vortices, instead of the correlation length,
will overestimate their number for 
$0.75\leq T \leq 0.85$, $\lambda = 0.0$ and 
$0.7\leq T \leq 0.8$, $\lambda = 0.9$.
In these temperature intervals, a vortex has its
nearest neighbor vortex at a distance smaller than $\xi(T)$, for
most of the vortices, and the resulting free vortex density will be
one or more orders of magnitude less than the theoretical
result Eq.~(\ref{nv}).
For $0.9\leq T \leq 1.1$, $\lambda = 0.0$ and
$0.85\leq T \leq 1.05$, $\lambda = 0.9$, 
the average number of free vortices should be approximately the same
when measured by using the correlation length or the $\sqrt{2}$
fixed-length scale.
The overestimation of the free vortex number very close to 
$T_{\mbox{\scriptsize KT}}$
will also lead to overestimation of the free vortex lifetime for these
values of T (see the next subsection).
However, this {\em does not change} the
conclusions regarding the magnitude of the lifetime with respect to the 
ideal vortex gas theory.
We also calculated the dependence of the average free and total vortex 
densities on $1/\xi_L^2$ to study their spatial distribution.
The results for the free and total vortex densities
are shown in Fig.~\ref{fig1} for 
$\lambda = 0.0$ and 0.9.
We do not obtain the straight line dependence of Eq.~(\ref{nv}) in the 
lower temperature range mainly because of the overestimation of the 
free vortex number in this range.
Nevertheless, the shape of these curves shows two slopes and tends to 
saturation in the upper temperature limit of these intervals, i.e. small 
$\xi$'s, which is more clearly seen if we include more data points at higher 
T's.
At these temperatures, the vortex motion is already diffusive and we are 
out of the range of ballistic motion assumed in the free-vortex gas theory.
Calculations of the vortex radial pair distribution functions 
shows that just above $T_{\mbox{\scriptsize KT}}$ the vortices are clustered
with very few isolated ones (the temperatures $T = 0.75,\ 0.8$ for 
$\lambda = 0.0$ and $T = 0.7,\ 0.75$ for $\lambda =0.9$).
Increasing the temperature decreases the number of vortices in the clusters
and for the smaller $\xi$'s in Figs.~\ref{fig1}, higher T's,
the vortices are approximately homogeneously distributed on the lattice.
This behavior is similarly observed in the plane
rotator model.\cite{TC79,Gup92}

\subsection{Vortex Lifetime}

\begin{figure}
\mbox{\hspace{.3in}\psfig{figure=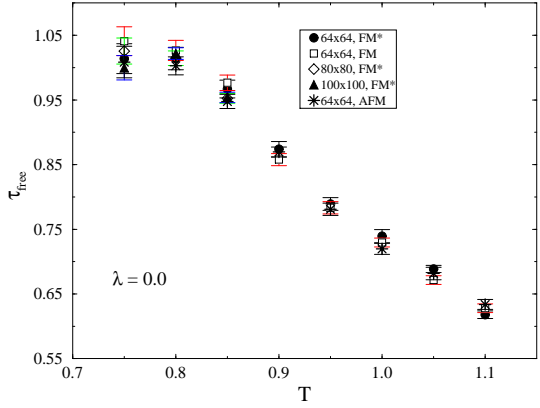,height=2.75in,angle=-90.}}
\caption{
  Free vortex lifetime vs temperature for $\lambda=0.0$, FM, AFM, 
  and several lattice sizes $L$. The FM* marked cases refer to 
  previous simulations using only Metropolis {\it et al.}\protect\cite{Met53}
  algorithm in the MC part of the simulations.}
\label{fig2} 
\end{figure}

\begin{figure}
\mbox{\hspace{.3in}\psfig{figure=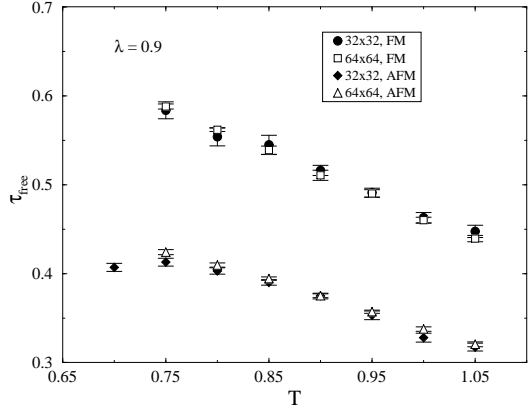,height=2.75in,angle=-90.}}
\caption{
  Free vortex lifetime vs temperature for $\lambda=0.9$, FM, AFM, 
  and the lattice sizes $L$ indicated.}
\label{fig3} 
\end{figure}

The free vortex lifetime is calculated
using a statistical method\cite{dad96}
during the spin dynamics part of the simulations.
The number of free vortices is counted at each time step $dt$ of the time
evolution and the times when this number decreases are saved in order to
determine the time intervals $\Delta t_i$ between
consecutive decreases of the number of free vortices.
Each of these intervals represents a data point for the calculation of the
free vortex lifetime from the following:
\begin{equation}
\label{taui} 
\tau_i ={ N_i \Delta t_i \over |\Delta N_i| }.
\end{equation}
The factor $N_i / |\Delta N_i|$ ($N_i$ is the number of free vortices
detected in the system just before the last time step and $|\Delta N_i|$
is the change in the free vortex number before and after the
last $dt$) accounts statistically for the possibility 
that any of the free vortices could have annihilated. 
The division by $|\Delta N_i|$ accounts for the case when
one time step evolution of
the system leads to a decrease of $N_i$ by more than one free vortex.
There are two requirements for the applicability of this method. 
The system has to be large enough in order to have a large number of free
vortices for better statistics when using Eq.~(\ref{taui}) and the time
step $dt$ has to be much smaller than the characteristic time of decay
$\Delta t_i$,  which helps to insure that in most cases $|\Delta N_i| = 1$.
The time step $dt$ has to be decreased when increasing $T$ or $L$ since then
the number of free vortices fluctuates on a shorter time scale.
The requirement to monitor the number of free vortices at each time step is 
the most CPU intensive task in the spin dynamics part of the simulations
and currently makes them excessively long if we employ fitting to
determine the precise vortex position as mentioned above.
We have used time steps in the interval
$1.5\times 10^{-4} \leq dt \leq 1.0\times 10^{-3}$ depending on $T$ 
and $L$ and trying to meet the condition $|\Delta N_i| \leq 2$.
Each Monte Carlo IC is evolved in time up to $t_{max} = 100$ by numerical
integration of the equations of motion, Eq.~(\ref{em}), using a
fourth-order Runge-Kuta scheme.
During the time evolution of the system,
$\tau_i$'s are calculated in order to determine an average lifetime for each
IC and its error.
The lifetimes for the different IC's and their errors are then additionally
averaged to obtain the final value for the lifetime and its error for given
temperature.
The results for FM and AFM and different lattice sizes $L$ are shown in
Fig.~\ref{fig2} for $\lambda = 0.0$
and in Fig.~\ref{fig3} for $\lambda = 0.9$.
The measured free vortex lifetimes for AFM and FM, $\lambda = 0.0$,
vary between $1.03$ for $T = 0.75$ and $0.63$ for $T = 1.1$.
Size effects are noticed for $L=32$, not plotted here,\cite{dad96} 
when $T=0.75$.
In this case the number of free vortices is very small and 
Eq.~(\ref{taui}) is not reliable for statistical analysis.
Note that the measured lifetimes for $0.75 \leq T \leq 0.85$ should
be considered as {\em an upper limit} for the free vortex lifetime in this 
temperature interval because of the overestimation of the number of free 
vortices, which leads to greater values of $N_i$ and thus of
$\tau_i$ calculated from Eq.~(\ref{taui}).
The free vortex lifetimes are the same, within the error bars, for FM
and AFM in the case of $\lambda = 0.0 $, Fig.~\ref{fig2}, regardless of the
different spin dynamics in both cases.
The picture is quite different for $\lambda > \lambda_c$ as we see in
Fig.~\ref{fig3} for $\lambda = 0.9$.
The observed lifetimes for AFM are smaller than those for FM and this can be
understood by the increased mobility of vortices and lower mass in AFM 
compared to FM.\cite{Wys96}
These measured lifetimes are to be compared with the characteristic time
scale Eq.~(\ref{char_time}) in the ideal vortex gas theory.\cite{Mer87}
These times are listed in Table~\ref{table3} for set of temperatures from the
published data on the correlation length and the vortex rms velocity obtained
from fitting the width and its integrated intensity of the observed
central peak in $S^{xx}({\bf q},\omega)$ in simulations.
The comparison of our results on the lifetime with the time scales in
Table~\ref{table3} shows that the lifetime is smaller than the characteristic
lifetimes as much as one order of magnitude for the higher temperatures
listed in Table~\ref{table3} while our results on the correlation length
for $\lambda = 0.0$ (Table~\ref{table1}) are between $\xi_1$ and $\xi_2$
(see Table~\ref{table3}) which are obtained from the central peak.
Even though the lifetimes we determined are shorter than the characteristic
times from the ideal-vortex-gas theory for the temperatures listed in
Table~\ref{table3}, we cannot rule out the validity of the theory to
describe the spin dynamics because it
does predict correlation lengths slightly smaller or larger, depending on
which quantity one fits, than the correlation length we determine
directly and one has yet to determine the rms vortex
velocity $\bar{u}$ independently.
We cannot currently measure $\bar{u}$ directly from the simulations but 
visualisation of the vortex positions at each time step shows that a free 
vortex almost never travels more than one lattice constant before
it either becomes bound by moving closer to other vortices, or another 
vortex moves close to it, or vortex-antivortex pair creation occurs close 
to it.
Our results on the vortex densities and clustering
for $T=T_{\mbox{\scriptsize KT}}+0$ and
the short vortex lifetime suggest that they may need be incorporated in the
theory rather than assuming infinite vortex lifetime.
Costa and Costa\cite{CC96} have stated that it may be necessary to 
consider other processes or theoretical descriptions for the cause of the 
CP in $S^{xx}({\bf q},\omega)$, in addition to the ideal gas theory.
In particular, they have suggested that vortex pair creation-annihilation
events could be the processes responsible the CP. 
Since there is presently no theory or phenomenology that leads to any 
quantitative or even qualitative predictions, we have no way to answer
this question for $T > T_{\mbox{\scriptsize KT}}$.
However, for $T \leq T_{\mbox{\scriptsize KT}}$ Evertz and Landau\cite{Eve96}
have made high-precision spin dynamics simulations for the FM model
with $\lambda=0$.
There they also found a CP in $S^{xx}({\bf q},\omega)$,
as well as other interesting unexplained features for frequencies
below the spin wave peak. 
This case is very interesting, because for 
$T \leq T_{\mbox{\scriptsize KT}}$ the free-vortex density is approximately
zero, whereas there can be a much larger bound vortex pair density.
Thus we can ask whether in this case pair creation and annihilation
events may be considered to cause the observed CP.
This question can be roughly answered by determining the vortex-antivortex
pair lifetime $\tau_{pair}$.
If annihilation-creation events are responsible for the observed CP,
then the width of the CP should be of the order of $\tau_{pair}^{-1}$.

Our method to determine the free vortex lifetime is directly applicable for
calculation of vortex-antivortex pair lifetime for
$T \leq T_{\mbox{\scriptsize KT}}$.
The pair lifetime is determined also from Eq.~(\ref{taui}) with
$N_i$ substituted by the total number of vortices and antivortices in the
system $N_{tot}$ and $|\Delta N_i|$ by $|\Delta N_{tot}|$.
Then $N_{tot}/|\Delta N_{tot}|$ will correctly be the number of
pairs before an
event of pair annihilation is observed, provided the time step $dt$ is small
enough and in a process of annihilation $N_{tot}$ decreases only by a
vortex-antivortex pair ($|\Delta N_{tot}| = 2$).
We calculated the pair lifetime at $T_{\mbox{\scriptsize KT}}$ ($T=0.7$) and
well below it at $T=0.4$ with $\lambda = 0.0$.
The lifetimes determined are $\tau_{pair} = 0.487(9)$ and $0.39(13)$,
respectively, with $L = 64$ and 128. 
The average vortex pair densities for these two simulations are 
$2.4(2)\times 10^{-5}$ for $T=0.4,\ L=128$, and $6.9(4)\times 10^{-3}$
for $T=0.7,\ L=64$.
The pair lifetime for $T=0.4$ may show size dependence because of the
the very low vortex pair density for lifetime estimation with
the linear lattice size equal to $L=128$. 
The time scale of these excitations gives approximately two orders of
magnitude higher frequency ($1/\tau_{pair}= 2.04$ and $2.56$ for
$T=0.7,\ 0.4$) than the observed frequency width,\cite{Eve96}
approximately 0.02 and $0.01$, of 
the central peak at these temperatures.
This rules out the explanation of the reported central peak for
$T\leq T_{\mbox{\scriptsize KT}}$ only by the vortex pair creation and
annihilation excitations.

%
%
\section{Conclusions}

We carried out combined cluster Monte Carlo and spin dynamics simulations on
classical two-dimensional easy-plane ferromagnets and antiferromagnets
for two values of the easy-plane strength.
The correlation length and vortex densities were calculated in the MC
simulations and their implications in the search for free vortices considered.
The lifetime of free vortices and vortex-antivortex pairs were calculated in
the spin dynamics simulations by averaging over the equilibrium spin
configurations supplied by the Monte Carlo runs.
The free vortex lifetime is of the same order of magnitude but smaller than
the characteristic time scale of the ideal vortex gas theory for
$\lambda = 0.0$ and $T_{\mbox{\scriptsize KT}} < T < 0.85$.
This result does not rule out the validity of the theory at least in
this interval before we have a direct way to determine the average
vortex velocity.
For higher temperatures and $\lambda = 0.9$, the lifetime becomes smaller
than the characteristic time by approximately one order of magnitude.
The free vortex lifetime in FM is greater than in AFM for $\lambda = 0.9$
most likely due to the greater vortex mass in FM compared to AFM and thus 
lower mobility.
For $\lambda = 0.0$, the lifetimes overlap for FM and AFM within the error
bars.
The vortex-antivortex pair lifetimes at $T = T_{\mbox{\scriptsize KT}}$ and 
$T = 0.4$ are approximately two orders of magnitude smaller than the
time scale of the observed central peak at
$T \leq T_{\mbox{\scriptsize KT}}$.
This suggests that the pair creation and annihilation excitations alone
can not be the reason for the central peak.

{\sl Acknowledgments.}---This work was supported by NSF Grants
DMR---9412300, CDA-9724289, 
and by NSF/CNPq International Grant INT---9502781. GMW also thanks FAPEMIG 
(Brazil) for a Grant for Visiting Researchers while at Universidade Federal
de Minas Gerais, Brazil, where a part of this work was performed.

%
%

%
%
%

%
%
\newpage
\begin{table}
\caption{Results for the correlation length $\xi_L$ for the
	temperatures, lattice sizes $L\times L$, and anisotropy
	parameter values $\lambda = 0,\ \ 0.9$ considered.
	The correlation length is in
	lattice constant units and temperature in $J/k_B$.}
\label{table1}
\begin{tabular}{r@{}lrr@{}l@{}c@{}@{}r@{}lr@{}l@{}c@{}@{}r@{}l}
 \multicolumn{2}{c}{Temp.}&\multicolumn{1}{c}{L}&
 \multicolumn{5}{c}{$\xi_L^{\lambda = 0.0}$}&
 \multicolumn{5}{c}{$\xi_L^{\lambda = 0.9}$}\\
\tableline
  0&.7   &256   &  &   &           &   &       &25&.88&{${}\pm{}$} & 0&.96 \\
   &     &128   &  &   &           &   &       &25&.48&{${}\pm{}$} & 0&.21 \\ 
  0&.75  &256   &51&.80&{${}\pm{}$}&  0&.5     &  &   &            &  &    \\
   &     &64    &  &   &           &   &       & 9&.03& {${}\pm{}$}& 0&.06 \\
  0&.8   &128   &11&.62&{${}\pm{}$}&  0&.36    &  &   &            &  &    \\
   &     &32    &  &   &           &   &       & 5&.15&{${}\pm{}$} & 0&.03 \\
  0&.85  & 64   & 6&.09&{${}\pm{}$}&  0&.20    &  &   &            &  &    \\
   &     & 32   & 6&.08&{${}\pm{}$}&  0&.02    & 3&.58&{${}\pm{}$} & 0&.03 \\
  0&.9   & 32   & 3&.95&{${}\pm{}$}&  0&.02    & 2&.77&{${}\pm{}$} & 0&.02 \\  
  0&.95  & 32   & 2&.93&{${}\pm{}$}&  0&.02    & 2&.29&{${}\pm{}$} & 0&.03 \\  
  1&.0   & 32   & 2&.35&{${}\pm{}$}&  0&.029   & 1&.88&{${}\pm{}$} & 0&.03 \\  
  1&.05  & 32   & 1&.95&{${}\pm{}$}&  0&.028   & 1&.69&{${}\pm{}$} & 0&.04 \\  
  1&.1   & 32   & 1&.63&{${}\pm{}$}&  0&.033   &  &   &            &  &    \\    
\end{tabular}
\end{table}

\begin{table}
\caption{
Characteristic times $1 / \gamma_i = 2\xi_i /
\protect\sqrt{\pi}\protect\bar{u}$,
$i=1,\,2$, in the ideal vortex gas theory from published data. The rms
vortex velocity $\bar{u}$ and the correlation length $\xi_2$ have been
obtained from fitting the width of $S^{xx}({\bf q},\omega)$, and $\xi_1$
from fitting the integrated intensity.\protect\cite{Mer87,Vol92b,Vol91}
}
\label{table3}
\begin{tabular}{ccr@{}lr@{}lcr@{}l}
Temp. & $\bar{u}$ & \multicolumn{2}{c}{$\xi_1$} &
   \multicolumn{2}{c}{$\xi_2$} &
   $1/\gamma_1$ & \multicolumn{2}{c}{$1/\gamma_2$}\\
\tableline
\multicolumn{9}{c}{FM, $\lambda = 0.0$, Ref.~\protect\onlinecite{Mer87}}\\
0.90 & 0.84 & 4&.4&  4&.8&  5.91 & 6&.45 \\
1.00 & 0.91 & 2&.4&  3&.0&  2.97 & 3&.72 \\
1.10 & 0.91 & 2&.1&  1&.9&  2.60 & 2&.36 \\
\multicolumn{9}{c}{AFM, $\lambda = 0.0$, Ref.~\protect\onlinecite{Vol91}}\\
0.85 & 1.17 & 4&.6&  9&.02& 4.44 & 8&.70 \\
0.90 & 0.96 & 3&.69& 5&.28& 4.34 & 6&.21 \\
0.95 & 1.05 & 2&.43& 4&.35& 2.61 & 4&.67 \\
1.00 & 1.05 & 2&.09& 3&.28& 2.24 & 3&.52 \\
1.05 & 1.13 & 1&.54& 3&.17& 1.53 & 3&.16 \\
\multicolumn{9}{c}{FM, $\lambda = 0.8$, Ref.~\protect\onlinecite{Vol92b}}\\
0.85 & 0.39 & 2&.80& 6&.60& 8.10 & 19&.09 \\
0.90 & 0.47 & 2&.03& 9&.53& 4.87 & 22&.88 \\
0.95 & 0.43 & 1&.72& 4&.32& 4.51 & 11&.33 \\
\multicolumn{9}{c}{AFM, $\lambda = 0.8$, Ref.~\protect\onlinecite{Vol92b}}\\
0.85 & 1.23 & 3&.29& 6&.87& 3.08 & 6&.30 \\ 
0.90 & 1.05 & 2&.25& 3&.74& 2.42 & 4&.01 \\ 
1.00 & 0.93 & 1&.56& 2&.26& 1.89 & 2&.74 \\ 
\end{tabular}
\end{table}

\end{document}